\documentclass{article} 

\usepackage{preprint}

\usepackage{amsmath, amsthm, amssymb, amsfonts}

\usepackage{svg}
\usepackage[numbers,square]{natbib}
\usepackage[utf8]{inputenc}	
\usepackage[T1]{fontenc}	
\usepackage{xcolor}		
\usepackage[colorlinks = true,
            linkcolor = purple,
            urlcolor  = blue,
            citecolor = cyan,
            anchorcolor = black]{hyperref}	
\usepackage{booktabs} 		
\usepackage{nicefrac}		
\usepackage{microtype}		
\usepackage{lineno}		
\usepackage{float}			

\usepackage{lipsum}		

\usepackage{newfloat}
\DeclareFloatingEnvironment[name={Supplementary Figure}]{suppfigure}
\usepackage{sidecap}
\sidecaptionvpos{figure}{c}

\usepackage{titlesec}
\titlespacing\section{0pt}{12pt plus 3pt minus 3pt}{1pt plus 1pt minus 1pt}
\titlespacing\subsection{0pt}{10pt plus 3pt minus 3pt}{1pt plus 1pt minus 1pt}
\titlespacing\subsubsection{0pt}{8pt plus 3pt minus 3pt}{1pt plus 1pt minus 1pt}

\usepackage{tikz,xcolor,hyperref}

\definecolor{lime}{HTML}{A6CE39}
\DeclareRobustCommand{\orcidicon}{
	\begin{tikzpicture}
	\draw[lime, fill=lime] (0,0) 
	circle [radius=0.16] 
	node[white] {{\fontfamily{qag}\selectfont \tiny ID}};
	\draw[white, fill=white] (-0.0625,0.095) 
	circle [radius=0.007];
	\end{tikzpicture}
	\hspace{-2mm}
}
\foreach \x in {A, ..., Z}{\expandafter\xdef\csname orcid\x\endcsname{\noexpand\href{https://orcid.org/\csname orcidauthor\x\endcsname}
			{\noexpand\orcidicon}}
}

\title{Adoption and Adaptation of CI/CD Practices in Very Small Software Development Entities: A Systematic Literature Review}

\usepackage{authblk}

\author[1\thanks{\tt{75148710@epg.unap.edu.pe}}]{Mario Ccallo-Luque\orcidA{}}
\author[1]{Alex Quispe-Quispe\orcidB{}}

\affil[1]{Escuela de posgrado de Informatica, National University of the Altiplano of Puno}
\affil[2]{Department of Biology, National University of the Altiplano of Puno}

\begin{document}

\twocolumn[ 
  \begin{@twocolumnfalse} 
  
\maketitle

\begin{abstract}
This study presents a systematic literature review on the adoption of Continuous Integration and Continuous Delivery (CI/CD) practices in Very Small Entities (VSEs) in software development. The research analyzes 13 selected studies to identify common CI/CD practices, characterize the specific limitations of VSEs, and explore strategies for adapting these practices to small-scale environments. The findings reveal that VSEs face significant challenges in implementing CI/CD due to resource constraints and complex tool ecosystems. However, the adoption of accessible tools like Jenkins and Docker, coupled with micro-pipeline practices and simplified frameworks such as ISO 29110, can effectively address these challenges. The study highlights the growing trend of microservices architecture adoption and the importance of tailoring CI/CD processes to VSE-specific needs. This research contributes to the understanding of how small software entities can leverage CI/CD practices to enhance their competitiveness and software quality, despite limited resources.
\end{abstract}
\keywords{Continuous Integration, Continuous Delivery, Very Small Entities, DevOps, Software Development}
\vspace{0.35cm}

  \end{@twocolumnfalse} 
] 



\section{Introduction}

Agility and speed of delivery are now critical factors for business success in today's software development landscape. High demand from markets for quality software, faster development time, and rapid responses to market changes have developed the call for continuous methodologies and practices that may allow optimization regarding software development processes \cite{9690862,CHEN201772}. Given this scenario, Continuous Integration (CI) and Continuous Deployment (CD) practices emerge as two basic cornerstones that allow the optimization of the software development life cycle \cite{humble2010continuous}.

CI/CD allows for a number of advantages, such as team interaction between different development teams, shorter development cycles, and the simplification of deployment processes \cite{fi14020063}. Nevertheless, the adoption of these practices faces problems, mainly by VSEs in the software development sector.

The integration of CI/CD into VSE teams demands special attention due to their unique particularities.VSEs are defined as organizations with a maximum of 25 employees and they represent a significant part of the software industry \cite{8399332}. While sharing many challenges with larger enterprises, VSEs work with more limited resources and face different time pressures \cite{Pardo_Ordoñez_2022}. Apart from limited knowledge and experience of CI/CD practices, cultural and organizational barriers are also factors that bring down the level of effective adoption in these entities.

This research explores in detail the applicability of CI/CD practices in the context of VSEs. The study tries to fill this knowledge gap with the help of a systematic literature review, whose execution plan is based on the guidelines for systematic reviews in software engineering proposed by Kitchenham and Charters \cite{kitchenham2007guidelines}.

This study aims to investigate the practices of Continuous Integration/Continuous Deployment comparatively and evaluate the suitability in software development VSEs. In this respect, the identification and classification of commonly used industrial CI/CD practices, the analysis of peculiar characteristics and limitations of VSEs with respect to the adoption of CI/CD practices, and the proposal for an initial conceptual framework for evaluating the applicability and adaptability of those practices in the VSE context will be performed. The organization of the paper is as follows: in Section 2, the detailed SLR methodology is presented; Section 3 reports the obtained results; Section 4 discusses the key findings, and hence their implication; and lastly, Section 5 concludes the study with a view to future research. With this systematic review, we intend to give an overview of how to integrate the CI/CD practices in VSEs, adding to the body of knowledge in software engineering for small-scale software development organizations.
\section{Background}
The evolution of software development involves fast and efficient delivery of high-quality products as a priority in its lifecycle. In this context, the importance of integrating Continuous Integration (CI) and Continuous Deployment (CD) practices has increased.
Continuous Integration, focuses on the automatic and frequent integration of code changes into a shared repository \cite{shahin2017continuous}. Continuous Deployment extends this concept, allowing fast and consistent delivery of software to end users \cite{7006384}. These practices are an integral part of the DevOps paradigm, which emerged between 2007 and 2010 as a response to the need to unify development (Dev) and operations (Ops) processes \cite{Zarour2019_fs}.

Very Small Entities or small software teams, characterized by teams with 25 employees or less, play a fundamental role in the software industry. They often focus on specific market areas or serve as vendors for larger companies \cite{8399332}. These teams are characterized by their agility and adaptability, but they also face unique challenges due to their limited resources and time constraints \cite{Pardo_Ordoñez_2022}.
To meet the current software development market requirements, CI/CD practices have gained relevance for VSEs \cite{9070849}, however, CI/CD implementation in VSEs faces special challenges due to limited resources, both financial and personnel, lack of technical knowledge, resistance to change in current development methods, and problems in automating testing and deployment \cite{ALSOLAI2020106214}.
This study aims to deepen the understanding of how VSEs can implement and benefit from CI/CD practices, taking into account their unique constraints and challenges. The purpose is to offer useful insights that can guide these entities in the successful implementation of CI/CD, thereby increasing their competitiveness and effectiveness in the software development market. This study addresses the particular needs of VSEs in terms of CI/CD, helping to fill a significant gap in the current literature on software development in resource-constrained environments.
\section{Methodology}
This study employs a Systematic Literature Review (SLR) methodology to investigate Continuous Integration/Continuous Deployment (CI/CD) practices and their applicability in Very Small Entities (VSEs) within software development. The SLR process adheres to the guidelines set forth by Kitchenham and Charters~\cite{kitchenham2007guidelines} for systematic reviews in software engineering, tailored to the specific context of CI/CD in VSEs.
\subsection{Research Questions}
The study addresses the following research questions:
\begin{enumerate} \item What are the most commonly used CI/CD practices in small software companies (VSEs)? \item What specific characteristics and limitations do VSEs face when implementing CI/CD? \item How can the identified CI/CD practices be adapted for application in the VSE context? \end{enumerate}
\subsection{Search Strategy}
We designed a comprehensive and targeted search strategy, following the recommendations of Garousi et al.~\cite{garousi2019guidelines} for conducting literature reviews in software engineering.
\subsubsection{Search Terms}
We constructed search terms using a combination of keywords related to CI/CD and VSEs:
\begin{itemize} \item CI/CD terms: 'Continuous Integration' OR 'Continuous Delivery' OR 'CI/CD' OR 'DevOps' \item VSE terms: 'Very Small Entities' OR 'VSE' OR 'Small Software Companies' OR 'Micro-enterprises' \end{itemize}
\subsubsection{Search String}
The resulting search string was:
(('Continuous Integration' OR 'Continuous Delivery' OR 'CI/CD' OR 'DevOps') AND ('Very Small Entities' OR 'VSE' OR 'Small Software Companies' OR 'Micro-enterprises'))
\subsubsection{Search Resources}
We selected the following electronic databases based on their relevance to software engineering and coverage of academic and professional literature:
\begin{itemize} \item IEEE Xplore Digital Library \item ACM Digital Library \item Scopus \item ScienceDirect \end{itemize}
\subsection{Study Selection Process}
\begin{figure}[H] \centering \includegraphics[width=0.5\textwidth]{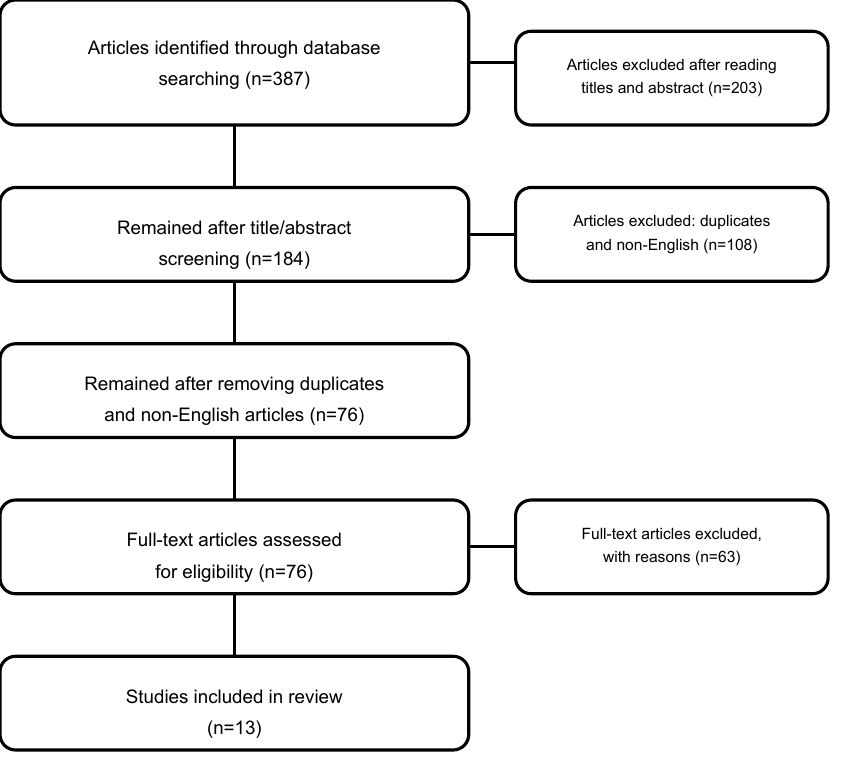} \caption{PRISMA flow diagram of the study selection process} \label{fig:prisma} \end{figure}

Our study selection process followed the PRISMA (Preferred Reporting Items for Systematic Reviews and Meta-Analyses) guidelines~\cite{MOHER2010336}. Figure~\ref{fig:prisma} illustrates this process.
Initially, we identified 387 articles from the selected databases. After reviewing titles and abstracts, we excluded 203 articles. Removing duplicates and non-English articles left us with 76. We then applied specific criteria, excluding articles that didn't align with our research questions. This rigorous process resulted in a final selection of 13 articles for detailed analysis.
\subsection{Inclusion and Exclusion Criteria}
We defined inclusion and exclusion criteria to ensure the relevance and quality of the selected studies.
\subsubsection{Inclusion Criteria}
\begin{itemize} \item Studies discussing CI/CD practices \item Studies addressing VSEs in the context of software development \item Peer-reviewed articles in journals and conferences \item Studies providing empirical or theoretical evidence on CI/CD application in VSEs \item English-language publications from the last 10 years (2014-2024) \end{itemize}
\subsubsection{Exclusion Criteria}
\begin{itemize} \item Studies unrelated to software development \item Grey literature (blogs, non-peer-reviewed technical reports) \item Studies lacking sufficient details on CI/CD practices or VSE characteristics \item Opinion articles, editorials, or conference abstracts \item Duplicate studies (we retained the most comprehensive version) \end{itemize}
\subsection{Data Extraction and Analysis}
We conducted data extraction using a predefined form to ensure consistency. Extracted data included study characteristics, methodologies used, key findings related to CI/CD practices in VSEs, and relevant insights addressing the research questions.
Our analysis of the extracted data followed a thematic synthesis approach. This involved coding the extracted data, identifying recurring themes, and synthesizing the findings to comprehensively address the research questions.
\subsection{Quality Assessment}
To ensure the reliability of our findings, we performed a quality assessment of the selected studies. This assessment considered factors such as the clarity of research objectives, the appropriateness of the methodology, the validity of data analysis, and the relevance of conclusions to the research questions.

\section{Results}

\begin{figure}[H]
    \centering
    \includegraphics[width=0.4\textwidth]{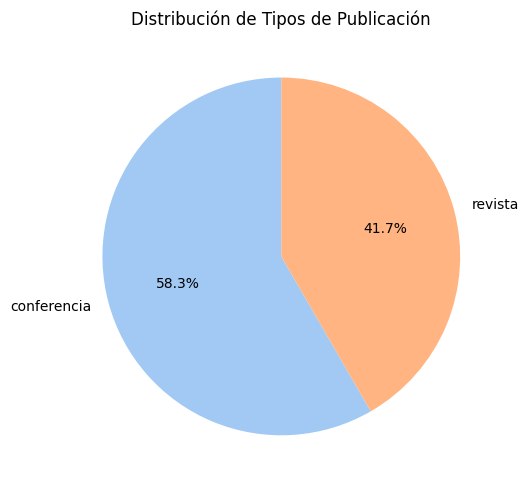}
    \caption{Distribution of publication types in the reviewed articles.}
    \label{fig:pub_dist}
\end{figure}

This section presents the insights obtained from reviewing 12 selected articles in our Systematic Literature Review (SLR). Our aim is to extract valuable information and results from the analysis and study of articles related to continuous integration (CI), continuous delivery (CD), and DevOps in small software enterprise (VSE) environments. In the analyzed dataset, comprising 12 extracted articles, a categorization emerges in terms of publication type, primarily differentiating between conference papers and journal articles. Of the reviewed articles, \textbf{83.3\%} correspond to conference papers, while \textbf{16.7\%} are classified as journal articles (see Figure \ref{fig:pub_dist}).

The articles are divided into two main types: \textit{conference papers} and \textit{journal articles}, reflecting a preference for conference publications within the study area of CI/CD in VSEs. This trend suggests that conferences are the preferred medium for disseminating applied research and case studies in this field.

\begin{figure}[H]
    \centering
    \includegraphics[width=0.5\textwidth]{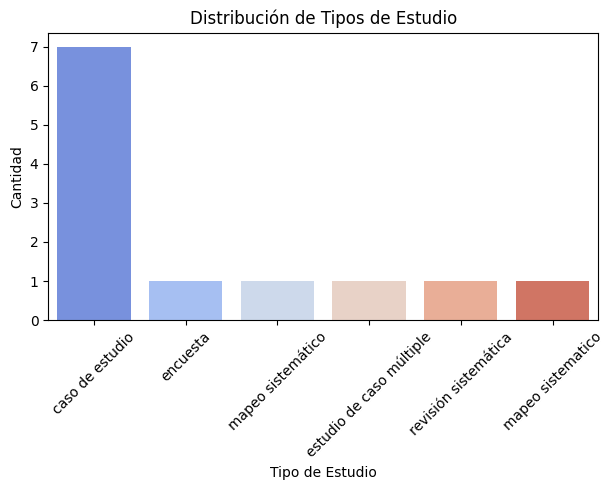}
    \caption{Distribution of study types in the reviewed articles.}
    \label{fig:study_dist}
\end{figure}

Of the 12 articles reviewed, the analysis distinguishes the following study types: a total of 7 articles (\textbf{58.3\%}) are classified as case studies, 3 articles (\textbf{25\%}) as systematic mapping studies, and 2 articles (\textbf{16.7\%}) as systematic literature reviews, as illustrated in Figure \ref{fig:study_dist}. These findings underscore the diverse research methodologies adopted in the literature, highlighting that the case study approach is predominant in research on CI/CD in VSEs.

Figure \ref{fig:pub_dist} presents the research design employed in the analyzed articles. Among these, seven cases are attributed to the case study type, while three are for systematic mapping studies, and two cases belong to the systematic literature review approach. This highlights the prevalence of empirical studies in the literature on CI/CD in the context of small enterprises.

\subsection{RQ1}

\begin{figure}[H]
    \centering
    \includegraphics[width=0.5\textwidth]{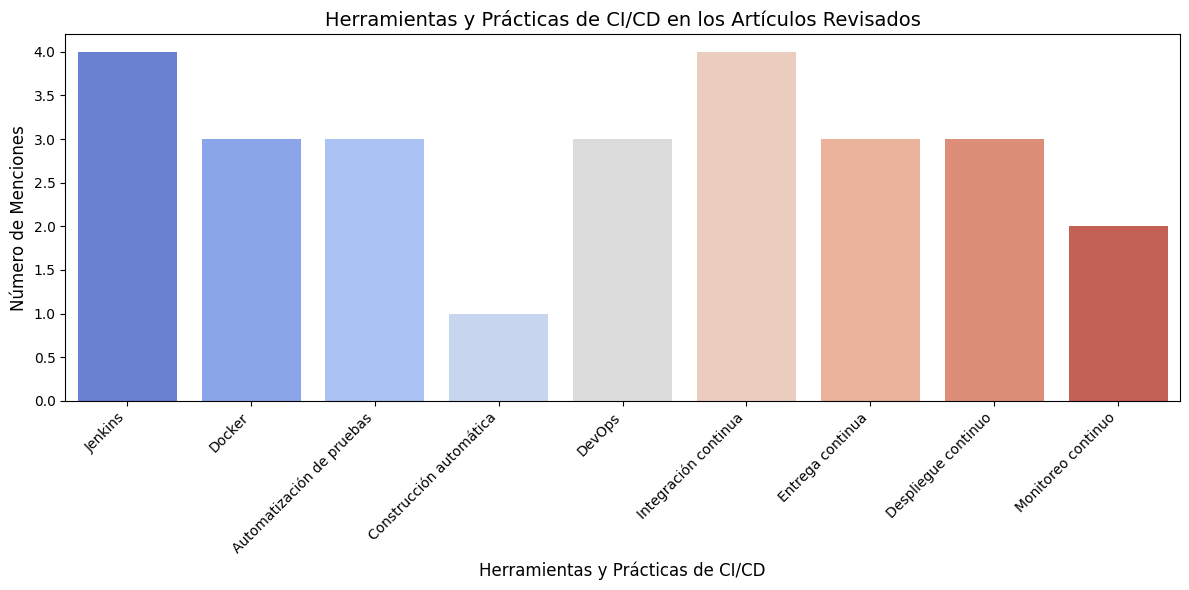}
    \caption{Most commonly used CI/CD practices in VSEs.}
    \label{fig:ci_cd_practices}
\end{figure}

The analysis of the 12 selected articles has allowed us to identify a series of key CI/CD practices adopted in small software companies (VSEs). These practices primarily focus on implementing accessible tools adapted to the resource and team size limitations. Below is a breakdown of the most commonly mentioned practices, illustrated in Figure \ref{fig:ci_cd_practices}.

Special importance is given to the frequency of the most used tools from VSEs for Continuous Integration and Continuous Deployment. The Figure shows that the most mentioned are Jenkins and Docker, in four and three articles respectively. Jenkins is described as the most used automation platform for handling continuous integration, and Docker is very frequently used alongside Jenkins.

On the other hand, the practices of continuous integration and continuous delivery as such are mentioned within DevOps, in four and three articles respectively. The other practices that appear are continuous deployment and monitoring. Three articles mention the practice of continuous deployment, where changes approved by the development team are automatically deployed to production. Meanwhile, the practice of continuous monitoring is mentioned in two articles.

\subsection{RQ2}

Very Small Entities (VSEs) face various specific characteristics and limitations when it comes to implementing Continuous Integration (CI) and Continuous Deployment (CD) practices.

\subsubsection{Characteristics and Limitations of VSEs in CI/CD Implementation}

VSEs, due to their reduced structure and limited resources, face significant challenges in implementing CI/CD. Studies such as those by Gunawan and Budiardjo \cite{10.1145/3451471.3451478} and Elazhary et al. \cite{9374092} show how resource constraints and time affect the ability of VSEs to implement efficient CI/CD practices, such as DevOps and Kanban, which often result in delays in software delivery. These challenges underscore the need to adopt specific frameworks and tools, such as ISO 29110 and Lean Six Sigma, to improve project management and development processes in these entities \cite{10.1145/3451471.3451478}.


\begin{figure}[htb]
    \centering
    \includegraphics[width=0.5\textwidth]{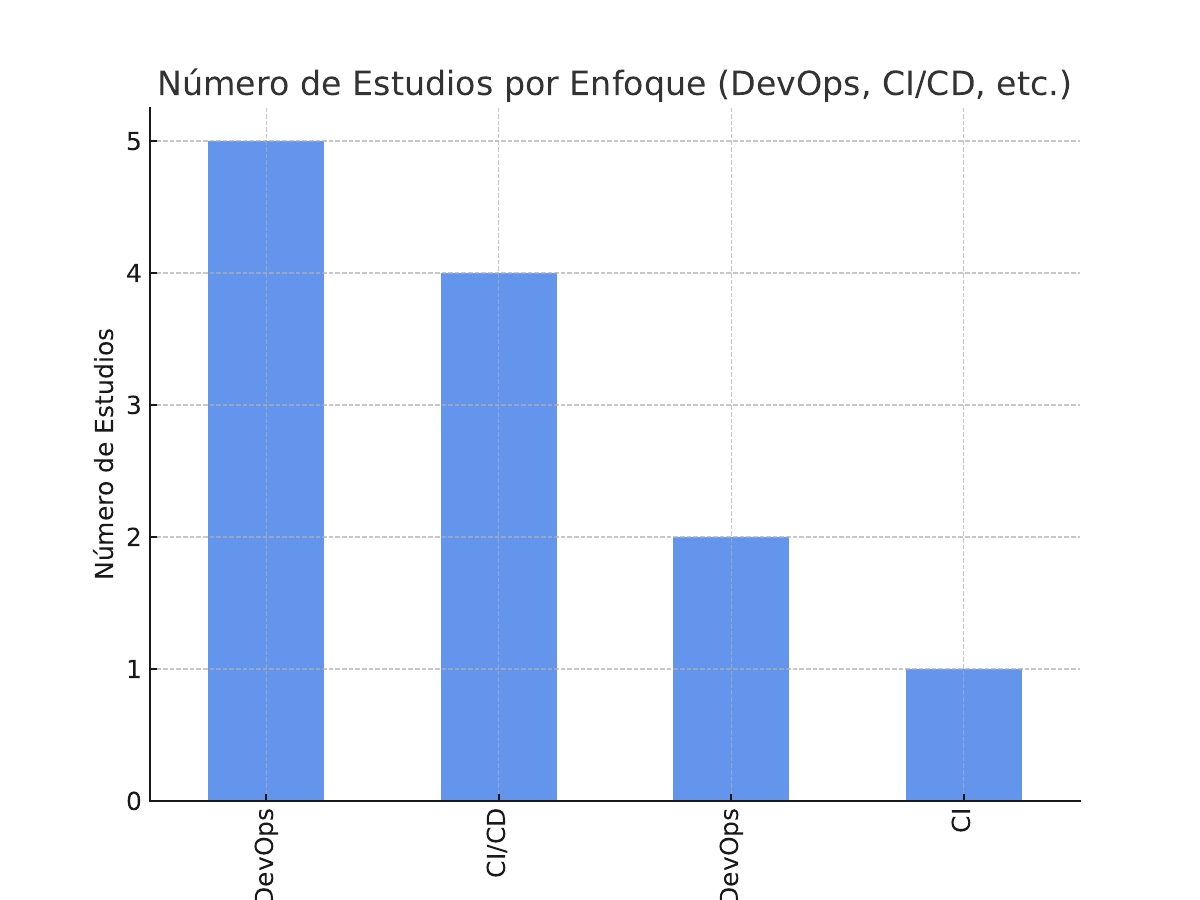}
    \caption{Number of Studies by Focus}
    \label{fig:focus_study}
\end{figure}

The distribution of studies by year figure \ref{fig:focus_study} shows how interest in research on CI/CD in VSEs has grown significantly between 2019 and 2023. This increase reflects a greater need for these entities to improve their processes to remain competitive in the software market. In particular, recent research underscores the importance of adapting advanced CI/CD practices to the scale and complexity of VSEs, which is crucial to overcoming resource management challenges \cite{8866741, 9070849}.

\begin{figure}[htb]
    \centering
    \includegraphics[width=0.5\textwidth]{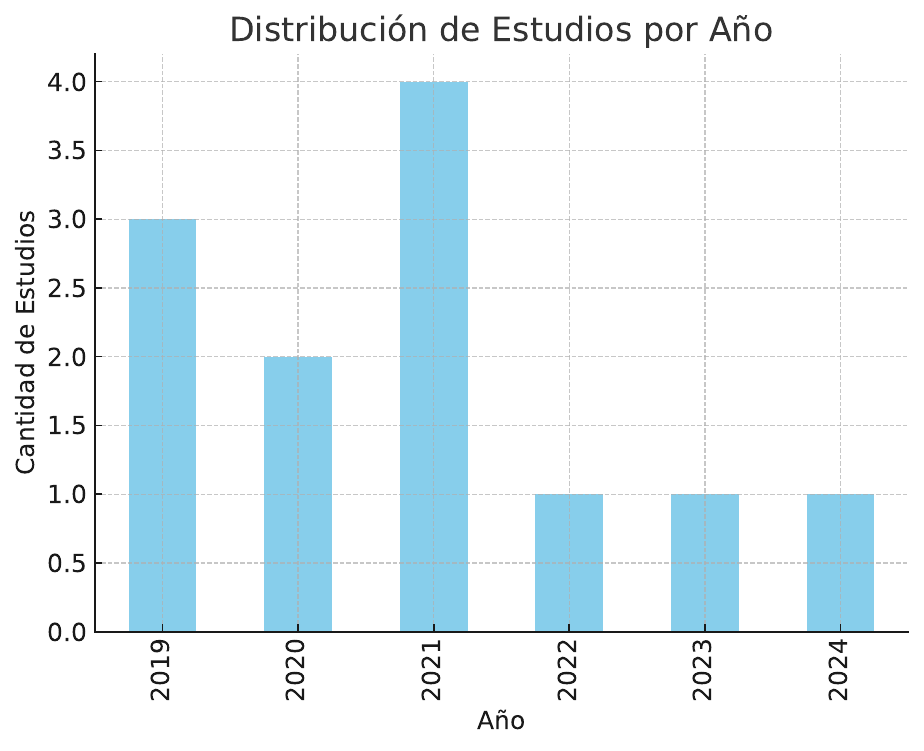}
    \caption{Distribution of Studies by Year}
    \label{fig:year_distribution}
\end{figure}

Another significant challenge is the adoption of CI/CD in microservices architectures. VSEs face difficulties in managing multiple dependencies and maintaining an efficient CI/CD infrastructure, especially when migrating to microservices-based architectures. Railić and Savić (2021) emphasize the importance of adopting advanced containerization and orchestration techniques, such as Docker and Jenkins, which allow managing independent pipelines for each microservice \cite{9400696}. Figure \ref{fig:year_distribution} supports this claim, showing an increase in interest in CI/CD as microservices architectures become more common in VSEs.

\subsubsection{Impact of Tool Complexity on VSEs}


Elazhary et al. \cite{9374092} and Tuape et al. \cite{10.1145/3501774.3501779} discuss how the use of complex tools also presents a significant barrier for VSEs. According to a study involving 115 participants from three countries, small software companies struggle with the complexity of requirements engineering and testing tools, which slows down the adoption of CI/CD tools \cite{10.1145/3501774.3501779, 10528811}. In these entities, the ability to adopt CI/CD frameworks that adapt to their limited resources and less formal organizational structures is crucial.

\begin{figure}[htb]
    \centering
    \includegraphics[width=0.5\textwidth]{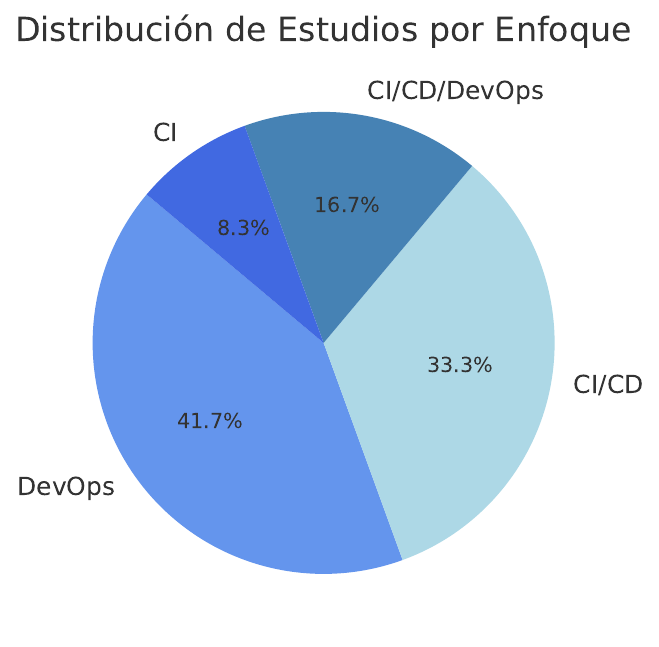}
    \caption{Distribution of Studies by Focus}
    \label{fig:focus_distribution}
\end{figure}

The distribution of studies by focus figure \ref{fig:focus_distribution} shows that much of the research has focused on DevOps and CI/CD, indicating that studies on the automation and simplification of these processes are vital for VSEs. To overcome these obstacles, the development of simplified platforms that allow small teams to implement CI/CD without facing a steep learning curve is recommended \cite{9070849, 10.1145/3451471.3451478}.

In this context, Abbass et al. (2019) proposed the use of micro-pipelines to optimize testing and reduce integration time, a strategy that has proven particularly useful in resource-constrained environments. The simplification of tools and processes, along with adequate training, becomes essential for VSEs to implement CI/CD efficiently \cite{9070849, 10.1145/3451471.3451478}.

\subsubsection{The Need for Standards and Implementation Guidelines}

The adoption of specific standards and guidelines is another critical need for VSEs. Research highlights the importance of developing adapted frameworks to facilitate the adoption of DevOps and CI/CD in these entities. Often, VSEs lack clear guidelines that allow them to effectively implement CI/CD tools and processes \cite{10528811, 8760865}. In this sense, alignment with standards such as ISO 29110 has shown to be a viable solution.

Previous studies demonstrated that VSEs can approach these standards using tools that are already familiar to them, which facilitates adoption and reduces resistance to change \cite{8760865, 9257512}. For this transition to be effective, it is essential that CI/CD tools and practices are specifically adapted to the needs of small businesses, thus avoiding an excessive learning curve and high implementation costs \cite{8866741, CASTILLOSALINAS2020103430}.

For example, Debroy and Miller (2020) analyzed how VSEs migrating from monolithic to microservices architectures face great difficulties in scaling their CI/CD pipelines. To mitigate these problems, they implemented a container-based and orchestration approach that allowed the company to efficiently manage deployments \cite{8866741}. Similarly, Lunde and Colomo-Palacios (2020) emphasize that continuous practices, such as DevOps, can help mitigate technical debt, which is essential for VSEs to better manage development complexity \cite{9257512}.

Finally, Muñoz et al. (2021) highlight how the adoption of a basic ISO/IEC 29110 profile to reinforce DevOps environments allows for a gradual and more effective implementation of maturity levels in automation, culture, and measurement \cite{muñoz2021platform}. Additionally, the study by Castillo-Salinas et al. (2020) demonstrates that even in resource-constrained contexts, the adoption of standards such as ISO/IEC 29110 significantly improves development processes, allowing VSEs to benefit from standardized practices without incurring high costs \cite{CASTILLOSALINAS2020103430}.

\subsection{RQ3}

Research question RQ3 seeks to adapt CI/CD practices for application in the context of Very Small Entities (VSEs), taking into account their specific characteristics and limitations. First, it is important to highlight that VSEs face substantial challenges when implementing CI/CD practices due to limited resources in terms of personnel, time, and budget. However, these practices are fundamental for improving software quality and accelerating development cycles, as evidenced by previous studies on CI/CD adoption in small organizations \cite{10.1145/3451471.3451478, 9070849}. Adaptations should focus on reducing the complexity of tools and processes used, favoring simplified platforms that can be managed by small teams without sacrificing efficiency \cite{8866741}.

One of the main strategies for adapting CI/CD in VSEs is the careful selection of tools that do not require an extensive learning curve or high implementation costs. Figures \ref{fig:focus_study}, \ref{fig:focus_distribution}, and \ref{fig:year_distribution} show that a significant portion of studies on CI/CD and DevOps in VSEs focuses on simplifying processes and implementing accessible solutions. Studies have revealed that the success of CI/CD implementation in VSEs largely depends on the ability to adapt the tools used to the characteristics of the entities, maintaining a flexible and pragmatic approach \cite{10528811, 9400696}.

Furthermore, another crucial aspect is the adoption of micro-pipeline practices, which allow VSEs to optimize the testing process in CI/CD. These practices divide the testing process into smaller, more manageable blocks, significantly reducing the time required for testing, allowing developers to make more frequent commits and detect errors more efficiently \cite{9070849}. This approach, based on optimizing the integration cycle, enables VSEs to overcome one of the main obstacles of traditional CI/CD: the prolonged duration of tests and the difficulty of managing multiple dependencies \cite{9374092}.

Finally, it is suggested that VSEs adopt architectures such as microservices to facilitate the scalability of their CI/CD processes. The adoption of a microservices-based architecture allows small entities to divide work into independent services that can be managed and deployed autonomously, reducing the overall complexity of the system \cite{8866741, 9400696}. Figures \ref{fig:year_distribution} and \ref{fig:focus_distribution} reinforce this idea, showing that the majority of studies related to CI/CD and DevOps in VSEs focus on automation and scalability, which is vital for improving their competitiveness in the software market.
\section{Conclusions}

This systematic review reveals that implementing CI/CD practices in Very Small Entities (VSEs) is a significant challenge, but not an insurmountable one. VSEs, due to their structural and resource limitations, face difficulties when attempting to adopt advanced software development and deployment practices, especially compared to larger organizations. However, the analyzed studies highlight a series of key strategies that can be used to mitigate these challenges and improve the competitiveness of VSEs in the software market.

First, the careful selection of accessible tools adapted to the needs of VSEs is fundamental. Jenkins and Docker emerge as the most commonly used platforms, due to their ability to automate integration and deployment processes at a relatively low cost and without requiring an excessive learning curve. This is crucial in an environment where teams are typically small and time is limited. The adoption of these tools, along with practices such as continuous integration and continuous delivery, allows VSEs to improve efficiency in software delivery without compromising quality.

Furthermore, process simplification is essential. The implementation of micro-pipelines and the adoption of agile methodologies such as DevOps and Kanban have proven effective in reducing complexity and accelerating development cycles in VSEs. This facilitates early error detection and reduces the time needed for testing, which is particularly valuable in resource-constrained contexts. Additionally, studies underline that the adoption of microservices architectures is a growing trend, allowing for greater flexibility and scalability in VSEs' CI/CD processes.

Another important conclusion is the need to adapt frameworks and standards to the specific characteristics of VSEs. The use of frameworks such as ISO 29110 and methodologies like Lean Six Sigma offer a more manageable and less costly structure for these entities. This not only reduces the complexity of CI/CD implementation but also allows them to align with globally accepted development practices without incurring high costs. In fact, previous studies have shown that the adoption of simplified standards improves software quality, even in contexts with restricted resources.

The research also highlights the importance of addressing the complexity of tools, which can be a significant barrier for VSEs. Developing simplified platforms that allow small teams to implement CI/CD without facing a steep learning curve is crucial. This includes optimizing testing processes and reducing integration time through strategies like micro-pipelines, which have shown to be particularly useful in resource-constrained environments.

Lastly, the study emphasizes the growing interest in CI/CD research for VSEs, particularly between 2019 and 2023. This trend reflects an increasing recognition of the importance of these practices for small software entities to remain competitive. The focus on DevOps and CI/CD in recent research underscores the critical nature of automating and simplifying these processes for VSEs.

\normalsize
\bibliography{references}

\end{document}